\documentclass[11pt]{article}
\newcommand{\Comment}[1]{{}}
\usepackage[textwidth = 430 pt, textheight = 630 pt]{geometry}
\usepackage{amssymb,euscript,amsmath,amsfonts}

\usepackage{color}
\definecolor{MyDarkBlue}{rgb}{0.15,0.15,0.45}
\usepackage[linktocpage=true]{hyperref}
\hypersetup{
colorlinks=true,
citecolor=MyDarkBlue,
linkcolor=MyDarkBlue,
urlcolor=MyDarkBlue,
pdfauthor={Neil Lambert, Constantinos Papageorgakis and Maximilian Schmidt-Sommerfeld},
pdftitle={M5-Branes, D4-Branes and Quantum 5D super-Yang-Mills},
pdfsubject={hep-th}
}

\usepackage[numbers,sort&compress]{natbib}
\usepackage{hypernat}

\newcommand\sfrac[2]{{\textstyle\frac{#1}{#2}}}
\newcommand\ignore[1]{}
\def\one{{\,\hbox{1\kern-.8mm l}}}

\newcommand{\tr}{\operatorname{tr}}

\newcommand{\Spin}{\operatorname{Spin}}

\newcommand{\SO}{\mathrm{SO}} 
\newcommand{\SU}{\mathrm{SU}} \newcommand{\U}{\mathrm{U}}

\newcommand{\Cset}{{\,\,{{{^{_{\pmb{\mid}}}}\kern-.45em{\mathrm C}}}}}

\newcommand{\nn}{\nonumber}
\newcommand{\ie}{{\it i.e.}}

\newcommand{\be}{\begin{equation}}
\newcommand{\ee}{\end{equation}}
\newcommand{\bea}{\begin{eqnarray}}
\newcommand{\eea}{\end{eqnarray}}

\begin{document}

\renewcommand{\thefootnote}{\fnsymbol{footnote}}

\rightline{CERN-PH-TH/2010-294}
\rightline{KCL-MTH-10-17}

   \vspace{1.8truecm}

\vspace{15pt}

\centerline{\LARGE \bf {\sc M5-Branes, D4-Branes}} \vspace{.5cm} \centerline{\LARGE \bf {\sc  and  Quantum 5D super-Yang-Mills }} \vspace{2truecm} \thispagestyle{empty} \centerline{
    {\large {\bf {\sc N.~Lambert${}^{\,a,}$}}}\footnote{On leave of absence from King's College London.}$^,$\footnote{E-mail address: \href{mailto:neil.lambert@cern.ch}{\tt neil.lambert@cern.ch}} {,}
    {\large {\bf{\sc C.~Papageorgakis${}^{\,b,}$}}}\footnote{E-mail address:
                                 \href{mailto:costis.papageorgakis@kcl.ac.uk}{\tt costis.papageorgakis@kcl.ac.uk}} {and} {\large {\bf{\sc M.~Schmidt-Sommerfeld${}^{\,a,}$}}}\footnote{E-mail address:
                                 \href{mailto:maximilian.schmidt-sommerfeld@cern.ch}{\tt maximilian.schmidt-sommerfeld@cern.ch} }
                                                           }

\vspace{1cm}
\centerline{${}^a${\it Theory Division, CERN}}
\centerline{{\it 1211 Geneva 23, Switzerland}}
\vspace{.8cm}
\centerline{${}^b${\it Department of Mathematics, King's College London}}
\centerline{{\it The Strand, London WC2R 2LS, UK}}

\vspace{2.0truecm}

\thispagestyle{empty}

\centerline{\sc Abstract}

\vspace{0.4truecm}
\begin{center}
\begin{minipage}[c]{380pt}{
    \noindent We revisit the relation of the six-dimensional $(2,0)$ M5-brane Conformal Field     Theory compactified on $S^1$ to 5D maximally supersymmetric Yang-Mills Gauge Theory.  We show     that in the broken phase 5D super-Yang-Mills contains a spectrum of soliton states that can be     identified with the complete Kaluza-Klein modes of an M2-brane ending on the M5-branes. This     provides evidence that the $(2,0)$ theory on $S^1$ is equivalent to 5D super-Yang-Mills with     no additional UV degrees of freedom, suggesting that the latter is in fact a well-defined     quantum theory and possibly finite.}
\end{minipage}
\end{center}

\vspace{.4truecm}

\noindent

\vspace{.5cm}

\setcounter{page}{0}

\newpage

\renewcommand{\thefootnote}{\arabic{footnote}}
\setcounter{footnote}{0}

\linespread{1.1}
\parskip 4pt

{}~
{}~

\makeatletter
\@addtoreset{equation}{section}
\makeatother
\renewcommand{\theequation}{\thesection.\arabic{equation}}

\section{Introduction}

Multiple M5-branes are believed to be described at low energies by a novel, interacting, strongly coupled, 6D CFT with $(2,0)$ supersymmetry. Very little is known about such a theory and it is not expected to have a Lagrangian description. According to the type IIA/M-theory duality it arises as the strong-coupling, UV fixed-point of multiple D4-branes whose dynamics are obtained from open string theory. At weak coupling and low energy the open string dynamics correspond to maximally supersymmetric Yang-Mills theory in 5D with gauge group $\U(N)$.

Thus there are three distinct theories at play. The first is the 6D $(2,0)$ supersymmetric M5-brane CFT. The second is the reduction of this theory to 5D dimensions, but including all the Kaluza-Klein (KK) modes.  This is the worldvolume theory for multiple D4-branes at any value of the coupling. Finally we can also consider, at least classically, 5D maximally supersymmetric Yang-Mills gauge theory. Since it is power-counting non-renormalisable it is not clear that there is a quantum theory that can be obtained from the classical Lagrangian without adding additional UV degrees of freedom.

On the other hand the D4-brane worldvolume theory includes a coupling of the form
\begin{equation}
S_{RR} \propto \int d^5x \;C_1\wedge {\rm tr}(F\wedge F)+\ldots\ ,
\end{equation}
We see that instanton states\footnote{These are BPS particles in 5D analogous to monopoles in 4D.} on the D4-brane worldvolume source the graviphoton $C_1$ of type IIA string theory. Furthermore the mass of such a particle state with instanton number $k$ in 5D Yang-Mills theory can be obtained from the BPS formula and is
\begin{equation}
M = |Z_5| + |Q_E|
\end{equation}
This matches nicely with the KK spectrum of a compactified 6D theory  \cite{Rozali:1997cb}. Thus, even in the Yang-Mills limit, we can identify a tower of states which seem to know about the extra direction of M-theory.

However there are  puzzles in the identification of KK states with instantons \cite{Berkooz:1997cq,Seiberg:1997ax}. Consider the unbroken phase where the D4-branes all lie on top of each other. Here the worldvolume theory does indeed possess smooth $\frac{1}{2}$-BPS particle soliton states where the gauge field is that of an instanton in the spatial directions. Since it breaks half of the 16 supersymmetries this leads to 8 Fermion zero modes and, according to the analysis of \cite{Jackiw:1975fn} (see also \cite{Osborn:1979tq}), these states are degenerate and form a $2^4=16$ dimensional set, \ie\ 8 Bosons and 8 Fermions. This indeed matches the 8 Bosonic and 8 Fermionic degrees of freedom arising from the M5-brane with one unit of KK momentum. However the instanton moduli space is $4Nk$-dimensional, where $k$ is the instanton number and $N$ the number of D4-branes, and includes some non-compact modes. In general, after quantisation, one expects that any non-compact directions in the moduli space will lead to a continuous spectrum of states. In this case $4$ of these modes can simply be thought of as the BPS particle's position in space and the associated continuous spectrum is just the spatial momentum.  However, even for $k=1,N=2$ one finds a non-compact scale size. This seems problematic since the reduction of the M5-brane theory at KK level one should not give rise to an additional continuous mode beyond the spatial momenta.

In addition 5D super-Yang-Mills is naively power-counting non-renormalisable and thus cannot be viewed as a quantum field theory in its own right. The resolution to these problems is generally thought to depend on a UV completion that will also introduce additional degrees of freedom, which will  contribute to the famous $N^3$ dependence of the free energy of $N$ M5-branes \cite{Klebanov:1996un}.

As is well-known, the dynamics of D-branes is determined by the quantisation of open strings that end on the D-brane. Despite recent progress the picture for M-branes is still unclear. However the most natural interpretation is that the dynamics of M5-branes is obtained by considering M2-branes that end on the M5-brane \cite{Strominger:1995ac}. The resulting solution for a single M5-brane is known as the self-dual string \cite{Howe:1997ue}. In the case of parallel M5-branes one expects to find smooth, finite energy string states corresponding to M2-branes stretched between any pair of M5-branes. The $(2,0)$ worldvolume theory of M5-branes is then thought to arise from the dynamics of these strings.\footnote{See also \cite{Witten:1995zh}.}

Upon compactification on $x^5$ these self-dual strings states lead to particle states (when the string wraps $x^5$) as well as string-like states (when the self-dual string does not wrap $x^5$). Because of self-duality these are not independent degrees of freedom. Rather one typically thinks of the particle states as giving rise to the perturbative states of five-dimensional super-Yang-Mills. The string states are then their electro-magnetic duals. However the M5-brane on a circle also admits a KK tower of massive states and as we have mentioned these should appear as solitonic states that carry a non-vanishing instanton number. Here we will attempt to show that five-dimensional super-Yang-Mills can correctly account for all the required KK modes without the need for additional degrees of freedom, although we will only consider the broken phase corresponding to separated branes. In particular we will see that the aforementioned problem regarding non-compact modes of the moduli space does not arise.\footnote{This was already   mentioned in \cite{Lambert:1999ua}.} We take the perspective that the unbroken phase is a singular point in the vacuum moduli space where the system is strongly coupled and difficult to analyse, even in an otherwise well-defined and UV-complete quantum field theory.

Our analysis can be taken to suggest that  no new degrees of freedom are required in order to identify 5D super-Yang-Mills as a 6D theory on $S^1$.  This implies  that 5D super-Yang-Mills is precisely the M5-brane 6D CFT on $S^1$ for any value of the radius. If so, since the M5-brane CFT is finite, it follows that 5D super-Yang-Mills is also well defined. Furthermore this suggests the natural conclusion that 5D super-Yang-Mills is in fact  UV-finite.\footnote{The claim that 5D super-Yang-Mills already contains all the states of the $(2,0)$ theory and is possibly finite is also made independently in \cite{Douglas:2010iu}, which we received after this work was substantially completed. Reference \cite{Douglas:2010iu} also includes arguments about the structure of divergences in 5D super-Yang-Mills.}

The rest of this paper is organised as follows: In Section~\ref{5D} we will review 5D super-Yang-Mills and its superalgebra. In particular we will show that the instanton number is identified with the momentum along the extra dimension. In Section~\ref{particle} we look for charged {\it particle} states in 5D super-Yang-Mills which carry instanton number, while in Section~\ref{string} we will find charged {\it string} states in 5D that carry instanton number. In Section~\ref{other} we consider some other states. In Section~\ref{Higgs} we look at the Higgs mechanism of the $(2,0)$-theory on $S^1$. Finally in Section~\ref{conclusion} we close with some comments.

\section{5D super-Yang-Mills}\label{5D}

Let us start by reviewing 5D maximally supersymmetric Yang-Mills theory. The field content consists of a vector $A_\mu$ with $\mu =0,1,2,3,4$, five scalars $X^I$ with $I=6,7,8,9,10$ and Fermions $\Psi$, all taking values in the Lie-algebra of the gauge group.
The supersymmetry transformations for five-dimensional super-Yang-Mills are
\begin{eqnarray}
\nonumber \delta_\epsilon X^I  &=& i\bar\epsilon \Gamma^I\Psi\\
\delta_\epsilon A_\mu &=& i\bar\epsilon\Gamma_\mu\Gamma_5\Psi\\
\nonumber \delta_\epsilon\Psi  &=&
\frac{1}{2}F_{\mu\nu}\Gamma^{\mu\nu}\Gamma_{5}\epsilon + D_\mu X^I\Gamma^\mu\Gamma^I\epsilon   - \frac{i}{2}[X^I,X^J]\Gamma^{IJ}\Gamma^5\epsilon
\end{eqnarray}
and the spinor $\epsilon$ satisfies $\Gamma_{012345}\epsilon=\epsilon$.
Here $D_\mu X^I = \partial_\mu X^I - i[A_\mu,X^I]$ and $F_{\mu\nu} = \partial_\mu A_\nu - \partial_\nu A_\mu - i [A_\mu,A_\nu]$. There is an invariant action given by
\begin{eqnarray}
\nonumber S &=& -\frac{1}{g^2_{YM}}\int d^5x\;\tr\Big(\frac{1}{4}F_{\mu\nu} F^{\mu\nu} + \frac{1}{2}D_\mu X^I D^\mu X^I -\frac{i}{2}\bar\Psi\Gamma^\mu D_\mu \Psi \\  &&\hskip3cm+ \frac{1
}{2}\bar\Psi \Gamma^5\Gamma^I[X^I,\Psi]- \frac{1}{4}\sum_{I,J}[X^I,X^J]^2\Big)\ .
\end{eqnarray}
Here we have taken all spinors to be those of eleven-dimensions (\ie\ real with 32 components) with $C = \Gamma_0$ as the charge conjugation matrix defined by $\Gamma_M^T = -C\Gamma_M C^{-1}$, $M=0,1,2,...,10$. We are using $x^5$ as the extra dimension associated with M-theory.

Let us compute the symmetry algebra of this theory. To begin with we consider the supercurrent:
\begin{eqnarray}
j^\mu =- \frac{1}{2g_{YM}^2}{\rm tr} \Big(\Gamma^\nu\Gamma^\mu\Gamma^I D_\nu X^I\Psi - \frac{1}{2}F_{\nu\lambda}\Gamma^{\nu\lambda}\Gamma^\mu\Gamma^5 \Psi + \frac{i}{2}\Gamma^\mu\Gamma^{IJ}\Gamma^5[X^I,X^J]\Psi\Big)\;,
\end{eqnarray}
which satisfies $\partial_\mu j^\mu=0$ on-shell.
We note that
\begin{eqnarray}
\bar\epsilon^\alpha\{Q_\alpha,Q_\beta\} = \int d^4x \bar\epsilon^\alpha \{Q_\alpha,j^0_\beta\} = \int d^4x \delta_\epsilon j^0\;.
\end{eqnarray}
If we write $\delta_\epsilon j^0_\alpha = (\delta_\epsilon j^0)_\alpha{}^\beta \epsilon_\beta$ then we deduce
\begin{eqnarray}
\{Q_\alpha,Q_\beta\} = -\int d^4x ((\delta_\epsilon j^0) C^{-1})^-_{\beta\alpha}\;,
\end{eqnarray}
where $((\delta_\epsilon j^0) C^{-1})^-  = \frac{1}{2}(1-\Gamma_{012345})(\delta_\epsilon j^0) C^{-1}$. A lengthy but straightforward calculation shows that
\begin{eqnarray}\nonumber
\{Q_\alpha,Q_\beta\}& =& P_\mu (\Gamma^\mu C^{-1})^-_{\alpha\beta} + Z_5 (\Gamma^5 C^{-1})^-_{\alpha\beta} +  Z_\mu^I (\Gamma^\mu\Gamma^I C^{-1})^-_{\alpha\beta} \\&& + Z_{5}^I (\Gamma^ 5\Gamma^I C^{-1})^-_{\alpha\beta}
+ Z_{\mu\nu\lambda}^{IJ}(\Gamma^{\mu\nu\lambda}\Gamma^{IJ}C^{-1})^-_{\alpha\beta}\;,
\end{eqnarray}
which for vanishing Fermions employs
\begin{eqnarray}
P_\mu &=&- \int d^4x T_{0\mu}\\
Z_5 &=&- \frac{1}{8g^2_{YM}}  \int d^4x \; {\rm tr}(F_{ij}F_{kl}\varepsilon_{ijkl})\\
 Z_5^I  &= &\frac{1}{g^2_{YM}} \int d^4x \; {\rm tr}(D_iX^I F_{0i} +i[X^I,X^J]D_0X^J)\\
 Z_ i^I &=&\frac{1}{2g^2_{YM}}  \int d^4x \partial_j{\rm tr}(X^IF_{kl}) \varepsilon_{ijkl}\\
 Z_{0}^I &=&\frac{1}{8g^2_{YM}} \int d^4x\; {\rm tr}([X^J,X^K][X^L,X^M]\varepsilon^{IJKLM})\\
 Z^{IJ}_{0ij} &=&\frac{1}{6g^2_{YM}} \int d^4x \partial_i{\rm tr}(X^ID_jX^J)\\
 Z^{IJ}_{ijk} &=&-\frac{i}{72g^2_{YM}} \int d^4x \partial_l {\rm tr}([X^K,X^L] X^M)\varepsilon_{ijkl}\varepsilon^{IJKLM}\;.
\end{eqnarray}
In the above $i,j,k=1,2,3,4$ and
\begin{equation}
T_{\mu\nu} =\frac{1}{g^2_{YM}}{\rm tr}\Big(F_{\mu\lambda}F_\nu{}^\lambda - \frac{1}{4}\eta_{\mu\nu}F^2 + D_\mu X^ID_\nu X^I - \frac{1}{2}\eta_{\mu\nu}D_\lambda X^I D^\lambda X^I  + \frac{1}{4}\eta_{\mu\nu}[X^I,X^J]^2 \Big)
\end{equation}
is the energy-momentum tensor. In particular we see that, in a broken vacuum where $\langle X^6\rangle\ne 0$, the central charge $Z^6_5$ can be identified with the electric charge of the unbroken gauge group (here we impose the Gauss-law constraint):
\begin{equation}
Q_E =Z_5^6 =   \frac{1}{g^2_{YM}}\oint \tr(\langle X^6\rangle F_{0i}) d^3S_i\
\end{equation}
and $Z^6_i$ with the dual magnetic charge of a string extended along $x^i$:
\begin{equation}
Q_{Mi} = Z_i^6 =  \frac{1}{2g^2_{YM}}\oint \tr(\langle X^6\rangle F_{kl} \varepsilon_{ijkl} )d^2S_j\ .
\end{equation}

It is helpful to compare this to the $(2,0)$ superalgebra that arises on the M5-brane. Its most general form  consistent with the projection $\Gamma_{012345}Q=-Q$ and symmetric under $\alpha \leftrightarrow\beta$ is\footnote{See {\it e.g.} \cite{Howe:1997et,D'Auria:2000ec}.}
\begin{eqnarray}
\{Q_\alpha,Q_\beta\}= P_m (\Gamma^m C^{-1})^-_{\alpha\beta}+  Z_m^I (\Gamma^m\Gamma^I C^{-1})^-_{\alpha\beta} + Z_{mnp}^{IJ}(\Gamma^{mnp}\Gamma^{IJ}C^{-1})^-_{\alpha\beta}\;,
\end{eqnarray}
where $m,n,p=0,...,5$ and $Z^{IJ}_{mnp}$ is self-dual in spacetime indices. Dimensionally reducing this algebra to five dimensions we see that we must identify $P_5$ with $Z_5$ and hence to the instanton number in five-dimensional super-Yang-Mills:
\begin{equation}\label{P5is}
P_5 =\frac{k}{R_{5}}=-\frac{1}{8g^2_{YM}}  \int d^4x \; {\rm tr}( F_{ij}F_{kl}\varepsilon_{ijkl})\;,
\end{equation}
where $i,j,k=1,2,3,4$. Since both the KK momentum and instanton number are quantised we must identify $R_5 = g^2_{YM}/4\pi^2$.
Note that $k\in \mathbb Z$ when tr is normalised to be the usual matrix trace in the fundamental representation, \ie ~${\rm tr} \one_{n\times n} = n$ for $\SU(n)$.
The central charges $Z^I_m$ with $m\ne 0$ ($Q_E$ and $Q_{Mi}$ in the 5D theory) are carried by self-dual strings extended along $x^m$, corresponding to an M2-brane along $x^m,x^I$ ending on the M5-branes \cite{Howe:1997ue}.

\section{Dyonic Instantons as Wrapped Strings with KK momentum}\label{particle}

Let us examine the BPS spectrum of 5D Super-Yang-Mills with non-vanishing instanton number. We start by looking at particle states in five dimensions. These correspond to self-dual strings that are wrapped around the M-theory circle. Throughout this paper we assume that the gauge group is $\SU(2)$, corresponding to two M5-branes.

We wish to look for time-independent (so that $D_0=0$) Bosonic  solutions and hence we let $F_{ij},F_{0i}$ and $X^I$ ($I=6,...,10$) be non-vanishing functions of $x^i$, where $i= 1,...,4$. We  select  the $X^6$ direction as special since we will separate the
D4-branes along $x^6$. We also reserve $x^5$ as the M-theory direction that is part of the M5 worldvolume. Following \cite{Lambert:1999ua} we can set the remaining scalars to zero and write the energy as
\begin{eqnarray}
  E  &=&\frac{1}{g^2_{YM}} \int d^4 x \;{\rm tr} \Big[ \frac{1}{4}F_{ij}F_{ij} + \frac{1}{2}F_{0i}F_{0i}+\frac{1}{2} D_0X^6D_0X^6+\frac{1}{2} D_iX^6D_iX^6 \Big]\\
\nonumber  &=& \frac{1}{g^2_{YM}} \int d^4 x\;{\rm tr} \Big[\frac{1}{8}(F_{ij} -\frac{1}{2}\varepsilon_{ijkl}F_{kl})^2
+\frac{1}{2} D_0X^6D_0X^6+ \frac{1}{2}(F_{0i}+D_iX^6)^2 \\
\nonumber && + \frac{1}{8}\varepsilon_{ijkl}F_{ij}F_{kl} - D_i X^6 F_{0i}\Big]\;,
\end{eqnarray}
where we have also fixed various possible choices of sign. The last two terms can be seen to be total derivatives. To minimise this bound we set
\begin{equation}\label{DI}
D_0X^6=0\;,\qquad F_{ij} = \frac{1}{2}\varepsilon_{ijkl} F_{kl}\;,\qquad D_iX^6 = - F_{0i} \ .
\end{equation}
Note that the Gauss law implies:
\begin{equation}
D_iD_i X^6=0\ .
\end{equation}
 In this case we see that
\begin{equation}
E \ge |P_5| + |Q_E|\ ,
\end{equation}
with equality iff (\ref{DI}) holds.

For solutions which satisfy (\ref{DI}) the YM supersymmetry transformations become
\begin{eqnarray}
\delta\Psi  &=& F_{0i}\Gamma_i(\Gamma_0 \Gamma_5 -\Gamma_{6})\epsilon +
\frac{1}{4}F_{ij}\Gamma_{ij}\Gamma_{5}(1-\Gamma_{1234})\epsilon\ .
\end{eqnarray}
Given that $\Gamma_{012345}\epsilon=\epsilon$ for an M5-brane, we see that the preserved supersymmetries satisfy $\Gamma_{056}\epsilon=\epsilon$ and $\Gamma_{05}\epsilon=\epsilon$. From the M5-brane perspective, with $x^5$ being the M-theory direction, these are the projectors associated to an M2-brane along $x^0,x^5,x^6$ (and hence an F-string along $x^0,x^6$) and a momentum wave along $x^{5}$.
 Let us look at solutions to these equations. The solutions for instantons are well known and smooth. One can also find smooth finite energy solutions to $D_iD_iX^6=0$. The case of a single $\SU(2)$ instanton was explicitly given in \cite{Lambert:1999ua} as
\begin{equation}
A_i = 2\frac{\rho^2}{ x^2(x^2+\rho^2)}\eta^a_{ij}x^j\frac{\sigma^a}{2}\;, \qquad X^6 = v\frac{x^2}{x^2+\rho^2}
\frac{\sigma^3}{2}\ ,
\end{equation}
where $\eta^a_{ij}$ are the 't Hooft matrices. It is important to see that the electric charge then reads
\be
Q_E = - 4\pi^2 v^2\rho^2g^{-2}_{YM}\;.
\ee
Thus $\rho$ must in fact be quantised and is no longer a modulus. Indeed by turning on an electric charge one introduces a potential on the moduli space $V  \propto \rho^2$ \cite{Lambert:1999ua}.  Upon quantisation the wavefunctions are those of a harmonic oscillator and thus there is a discrete spectrum (with the exception of the continuous momentum modes).

We should compare this against the predictions of the $(2,0)$ theory.
Let us look for BPS states with mass $M$ and momentum $P_5$ along $x^{5}$.  Upon reduction to 5D the electric-central charge of the Yang-Mills theory comes from $Z^{6}_{5}$. The other central charges are not carried by particle states. Thus we only consider $Z^{6}_{5}\ne 0$.
In this case  we find
\begin{equation}
\{Q_{\alpha},Q_{\beta}\} = (M+P_5\Gamma^{50}+Z^6_5\Gamma^5\Gamma^6\Gamma^0)^-_{\alpha\beta }\;.
\end{equation}
To look for BPS states we need to find zero-eigenstates of $M +P_5\Gamma_{05} -Z^6_5\Gamma_{05}\Gamma^6$. Since $[\Gamma_{05},\Gamma_{05}\Gamma^6]=0$  these are simultaneous eigenstates of $\Gamma_{05}$ and $\Gamma_{05}\Gamma^6$. Thus we see that only a quarter of the supersymmetries survive. This is in agreement with the dyonic instantons. In addition we recover the same BPS formula for the mass.

Since the dyonic instanton has 12 Fermionic zero-modes it will come in a degenerate multiplet with
$64$ states \cite{Jackiw:1975fn}.  Let us examine this multiplet from the point of view of massive particles in 5D.  The 16 supersymmetries of the M5-brane can be labelled as $Q_{\eta_1,\eta_2}$ where $\eta_{1/2}=\pm$ and
\begin{equation}
\Gamma _{05}Q_{\eta_1,\eta_2}=\eta_1Q_{\eta_1,\eta_2}\;,\qquad \Gamma_{05}\Gamma^6Q_{\eta_1,\eta_2}=\eta_2Q_{\eta_1,\eta_2}\;,\qquad \Gamma_{012345}Q_{\eta_1,\eta_2}=-Q_{\eta_1,\eta_2}
\end{equation}
and hence $\Gamma^6Q_{\eta_1,\eta_2}=\eta_1\eta_2Q_{\eta_1,\eta_2}$.
The preserved supersymmetries of the dyonic instanton can be taken to be $Q_{-+}$ and therefore the broken supersymmetries are $Q_{--},Q_{++}$ and $Q_{+-}$. Each of these has 4 real components. For $Z^6_5\ne 0$ the R-symmetry is broken from $\mathrm{Spin}(5)$ to $\mathrm{Spin}(4)\simeq\SU(2)\times \SU(2)$. Since  $\Gamma^{678910}Q_{\eta_1,\eta_2}=-Q_{\eta_1,\eta_2}$ the broken supersymmetries are chiral spinors under this  $\SU(2)\times \SU(2)$. In particular the four components of  each of $Q_{--}$ and $Q_{++}$ can be arranged into complex generators $\bf Q_{--}$  and $\bf Q_{++}$, each of which is  in the $(\bf 2,\bf 1)$ of $\SU(2)\times \SU(2)$. Similarly the four components of  $Q_{+-}$ can be arranged as a complex $\bf Q_{+-}$ transforming as  $(\bf 1,\bf 2)$ under $\SU(2)\times \SU(2)$.

In five dimensions the little group of a massive state is $Spin(4)\simeq\SU(2)\times \SU(2)$, \ie\ there is a second $\SU(2)\times \SU(2)$. Each of the ${\bf Q}_{\eta_1,\eta_2}$ is also a chiral (or anti-chiral) spinor of this $\SU(2)\times \SU(2)$. In particular $\bf Q_{++}$ and $\bf Q_{+-}$ transform under the first $\SU(2)$ and  $\bf Q_{--}$ transforms under the second $\SU(2)$.\footnote{Note that these $\SU(2)$'s do not act on ${\bf Q}_{\eta_1,\eta_2}$ through the standard way and in particular rotate ${\bf Q}_{\eta_1,\eta_2}$ into ${\bf Q}^\star_{\eta_1,\eta_2}$.}  Since $\{{\bf Q},{\bf Q}^\dag\} = 1$ (where we have suppressed indices) we can use ${\bf Q}_{\eta_1 ,\eta_2}$ as lowering operators and ${\bf Q}^\dag_{\eta_1 ,\eta_2}$ as raising operators. We assume that there is a highest weight state $|s_1,s_2\rangle$ that is annihilated by all the $\bf Q_{\eta_1,\eta_2}^\dag$ and is a singlet of the R-symmetry group.  Acting with ${\bf Q}_{+,+}$ or ${\bf Q}_{+,-}$ lowers the $s_1$ weight by $\frac{1}{2}$ whereas acting with $\bf Q_{--}$ lowers the $s_2$ weight by $\frac{1}{2}$. Thus the lowest weight state is $|s_1-2,s_2-1\rangle$. Since the spectrum must be CPT self-conjugate we require that $|s_1-2,s_2-1\rangle = |-s_1,-s_2\rangle$ and hence the highest weight state is $|1,\frac{1}{2}\rangle$. We obtain the remaining states by acting with various powers of ${\bf Q}_{++},{\bf Q}_{+-}$ and ${\bf Q}_{--}$ to find
\begin{eqnarray}
\nonumber &&|1,\sfrac{1}{2}\rangle\  ({\bf 1},{\bf 1})\\
\nonumber &&|1,0\rangle\  ({\bf 2}, {\bf 1})\qquad |\sfrac{1}{2},\sfrac{1}{2}\rangle\ ({\bf 1} ,{\bf 2})\oplus ({\bf 2},{\bf 1})\\
&&|\sfrac{1}{2},0\rangle\  ({\bf 2}, {\bf 2})\oplus ({\bf 1}, {\bf 1})\oplus ({\bf 3}, {\bf 1})\qquad |0,\sfrac{1}{2}\rangle\  ({\bf 1},{\bf 1})\oplus({\bf 1},{\bf 1})\oplus({\bf 2} ,{\bf 2})\\
\nonumber&&|0,0\rangle\ ({\bf 3},{\bf 2})\oplus ({\bf 2},{\bf 1})\oplus ({\bf 2},{\bf 1})\oplus ({\bf 1},{\bf 2}) \\
\nonumber&& \vdots
\end{eqnarray}
where we have given their R-symmetry representations and the ellipses denote  states $|s_1,s_2\rangle $  with $s_1<0$ or $s_2<0$ which have the same representation as the corresponding state $||s_1|,|s_2|\rangle$.

Thus the field content of the multiplet (in the massive rest frame) is as follows.  For the Fermions we find a chiral $\psi_{ij}^+$ $(\bf 1,\bf 1)$ which is self-dual in the spacetime  $i,j$ indices, 5 chiral   $\lambda $ $(({\bf 1},{\bf 1})\oplus({\bf 2} ,{\bf 2}))$ and 8 anti-chiral   $\chi$ $ (({\bf 1}, {\bf 1})\oplus({\bf 2}, {\bf 2})\oplus  ({\bf 3}, {\bf 1}))$. For the Bosons  we find a complex doublet of self-dual 2-forms $B_{ij}^+$ $({\bf 2} ,{\bf 1})$,  four complex vectors $A_i$   $(\bf1 ,\bf 2)\oplus (\bf 2,\bf 1)$ and 10 scalars $\phi$ $ (({\bf 3},{\bf 2})\oplus ({\bf 2},{\bf 1})\oplus ({\bf 1},{\bf 2}))$. We note that these form $64$ complex states. This differs from the 64 real states one might have expected since in the complete CPT-symmetric multiplet we should also find the states with opposite charges, obtained from dyonic instantons with the opposite choice of charge ($F_{0i}= D_iX^6$). We also note that since the highest weight state is a Fermion, the Bosons transform under spinor representations of the $\SO(4)$ R-symmetry which makes any spacetime interpretation obscure.

The quantum mechanical treatment of the dyonic instanton moduli space also admits a normalisable ground-state wavefunction  corresponding to  a $\frac{1}{2}$-BPS instanton solution without charge. The expectation value of the size is $\langle \rho\rangle \propto \sqrt{g^2_{YM}/v}$. Thus in the $v>0$ phase it would appear that quantum effects lead to a non-singular uncharged $\frac{1}{2}$-BPS instanton solution with a finite, fixed value of $\rho$.

Once again we can compare these states to the predictions of the $(2,0)$ theory. In this case we need to consider zeroes of
 \begin{equation}
\{Q_{\alpha},Q_{\beta}\} = (M+P_5\Gamma^{50})^-_{\alpha\beta }\;.
\end{equation}
Clearly these preserve the  $Q_{--}$ and $Q_{-+}$ supersymmetries, which can be used to compute the multiplet of these uncharged instantons, in close analogy to the dyonic instantons above. The R-symmetry is still broken to $\SO(4)$, while the broken supersymmetries $Q_{++}$ and $Q_{+-}$ are chiral with respect to the $\Spin(4)\simeq \SU(2) \times \SU(2)$ massive little Lorentz group. We then see that the CPT self-conjugate representation has highest weight $|1,0\rangle$. Acting with the lowering operators produces
\begin{eqnarray}\label{tensormult}
\nonumber &&|1,0\rangle\  ({\bf 1},{\bf 1})\\
\nonumber &&|\sfrac{1}{2},0\rangle\  ({\bf 2}, {\bf 1})\oplus ({\bf 1} ,{\bf 2}) \\
&&|0,0\rangle\  ({\bf 2}, {\bf 2})\oplus ({\bf 1}, {\bf 1})\oplus ({\bf 1}, {\bf 1})\\
\nonumber&& \vdots
\end{eqnarray}
where again the ellipses denote states with  $s_1<0$ which have the same representations as the corresponding states with $s_1>0$. Thus we find a multiplet consisting of a self-dual tensor $B_{ij}^+$, four chiral Fermions $\lambda $ $({\bf 2}, {\bf 1})\oplus ({\bf 1} ,{\bf 2})$  and five scalars in the $({\bf 2}, {\bf 2})\oplus ({\bf 1}, {\bf 1})$ of the R-symmetry. These states are therefore in agreement with the KK tower of instanton states discussed in \cite{Hull:2000cf}.

However, we also note that the one-instanton moduli space has a factor ${\mathbb R}^+$ from the scale size and a factor of $\SU(2)$ from the embedding of the solution into the gauge group ({\it i.e.} $A_i\to U A_i U^\dag$ with $U\in \SU(2)$) - as well as a trivial factor of ${\mathbb R}^4$ from translations. Since this $\SU(2)\equiv S^3$ acts in the adjoint, the moduli space is really $\SO(3)\equiv S^3/{\mathbb   Z}_2$. Thus the (non-translational part) of the moduli space is singular: ${\mathbb   R}^4/{\mathbb Z}_2\equiv {\mathbb R}^+\times S^3/{\mathbb Z}_2$. Therefore to define the quantum mechanics of the moduli space might  involve additional  subtleties.

\section{Monopoles as Unwrapped Strings with KK momentum}\label{string}

Having obtained dyonic instanton solutions which describe the KK tower of wrapped self-dual strings we need to look for KK towers of five-dimensional strings, corresponding to M2-branes that do not wrap the $x^5$ direction. In particular, the D4-brane theory has $\frac{1}{2}$-BPS string states, corresponding to D2-branes ending on a D4. From the M5-brane perspective these states should therefore also allow for generalisations that carry KK momentum and hence instanton number.

 An infinitely long string moving in six dimensions will carry infinite momentum and this would therefore translate into infinite instanton number.  However we can regulate this by assuming that the string is wrapped along the $x^4$ direction and that $x^4$ is periodic with period $2\pi R_4$.
We then need to look for states which are invariant under translations along $x^4$ but carry instanton number
\begin{equation}
-k= \frac{1}{32\pi^2}\int d^4 x \; {\rm tr}(F_{ij}F_{kl}\varepsilon_{ijkl})  = \frac{R_4}{4\pi} \int d^3x \; {\rm tr}(F_{ab}D_cA_4\varepsilon_{abc})  =  \frac{R_4}{4\pi}\oint d^2 S_{c}\; {\rm tr}(F_{ab} A_4\varepsilon_{abc})\;,
\end{equation}
where  $a,b,c=1,2,3$ and $d^2S_c$ is the measure on the transverse two-sphere at infinity. In addition,  in five dimensions strings are the electromagnetic duals to charged particles and therefore have magnetic charges. Thus we require that
\begin{eqnarray}
Q_{M4} = \frac{1}{2g^2_{YM}}\oint d^2S_{c}\; {\rm tr} (F_{ab} X^6 \varepsilon_{abc})
\end{eqnarray}
is non-zero and we need to consider a configuration with $F_{ab}$, $X^6$ and $A_4$ non-vanishing.

As a result we look for static solutions with $D_0A_4=D_0X^6=F_{0a}=F_{04}=0$ . Let us repeat the Bogomoln'yi argument used above for this case:
\begin{eqnarray}
  E  &=&\frac{2\pi R_4}{g^2_{YM}} \int d^3 x \;{\rm tr} \Big[ \frac{1}{4}F_{ab}F_{ab}+  \frac{1}{2} D_aA_4D_aA_4+\frac{1}{2} D_aX^6D_aX^6  - \frac{1}{2} [A_4,X^6]^2 \Big]\\
\nonumber  &=& \frac{2\pi R_4}{g^2_{YM}} \int d^3 x\;{\rm tr} \Big[\frac{1}{4}(\sin\theta F_{ab} -\varepsilon_{abc}D_cA_4)^2+\frac{1}{4}(\cos\theta F_{ab} -\varepsilon_{abc}D_cX^6)^2 +\frac{1}{2} (i[A_4,X^6])^2\\
\nonumber &&
  + \frac{1}{2}\varepsilon_{abc}F_{ab}D_c(\sin\theta A_4 + \cos\theta X^6) \Big]\;,
\end{eqnarray}
where $\theta$ is an arbitrary angle. The last term is a total derivative so that the BPS equations are
\begin{eqnarray}\label{monopole}
F_{ab} = \varepsilon_{abc} D_c\Phi\;, \qquad A_4 = \sin\theta \Phi\;,\qquad X^6 = \cos\theta \Phi
\end{eqnarray}
and the Bianchi identity implies $D^2 \Phi=0$. Similar states with a six-dimensional interpretation were considered in \cite{BlancoPillado:2006qu}.
These are nothing more than the equations for a $\frac{1}{2}$-BPS monopole with scalar $\Phi$ that is a linear combination of $X^6$ and $A_4$. At large distances from the string, the solution to \eqref{monopole} behaves as\footnote{We remind that we work with an $\SU(2)$ gauge group.}
\begin{equation}
\Phi = \phi_0 \sigma^3 - \frac{q\sigma^3}{ 4\pi r}+\ldots\;,
\end{equation}
where $\phi_0$ is arbitrary and $q\in{\mathbb Z}$ is the monopole charge, in the sense that $\frac{1}{2}\oint d^2S_{c}F_{ab} \varepsilon_{abc}= q\sigma^3$. In particular, if we take
\begin{equation}
\cos\theta  = \frac{v/2}{\sqrt{v^2/4 +k^2 \pi^2/R_4^2q^2}}\;,\quad \sin\theta = \frac{k\pi/R_4q}{\sqrt{v^2/4 + k^2 \pi^2/R_4^2q^2}}\;,\quad \phi_0 = \sqrt{v^2/4 + k^2 \pi^2/R_4^2q^2}
\end{equation}
we then  find solutions with
\begin{equation}
P_5 = -\frac{ 4\pi^2 k}{g^2_{YM}}\;,\qquad Q_{M4}  = \frac{vq}{g^2_{YM}}\;,
\end{equation}
 as required, the energy of which is given by
\begin{equation}
E =  {2\pi R_4  } \sqrt{Q_{M4}^2+(P_5/2\pi R_4)^2}\;.
\end{equation}
The one-monopole moduli space  is well-studied and only has 3 non-compact modes corresponding to translations transverse to the string in five-dimensions. Thus there is no issue with obtaining a continuous spectrum of states.

For completeness, let us check the amount of supersymmetry preserved by  these solutions. We see that:
\begin{eqnarray}
\nonumber 0&=&\delta \psi\\
 &=&\nn\frac{1}{2}F_{ab}\Gamma^{ab}\Gamma^5\epsilon + D_aA_4\Gamma^{a4}\Gamma^5\epsilon+ D_aX^6\Gamma^a\Gamma^6\epsilon-i[A_4,X^6]\Gamma^4\Gamma^6\epsilon\\
 \nonumber &=& (\frac{1}{2}\epsilon_{abc}\Gamma^{ab}\Gamma^5 + \Gamma^c(\sin\theta\Gamma^4\Gamma^5+\cos\theta\Gamma^6))D_c\Phi\epsilon\\
  &=&  \Gamma_c\Gamma_0\Gamma_4(1 -\sin\theta\Gamma^0\Gamma^5+\cos\theta\Gamma_0\Gamma^{46})D_c\Phi\epsilon\;,
\end{eqnarray}
which can be solved by the $\frac{1}{2}$-BPS projection $(\sin\theta\Gamma^0\Gamma^5-\cos\theta\Gamma_0\Gamma^{46})\epsilon=\epsilon$.

We should also compare this with the 6D $(2,0)$ predictions for a state with central charge $Z^6_{4}$ and momentum $P_5$. In particular we now find
\begin{equation}
\{Q_{\alpha },Q_{\beta }\} = (M+P_5\Gamma_{05} +Z^6_4\Gamma^{4}\Gamma^6\Gamma^0)^-_{\alpha\beta}\;.
\end{equation}
Just as in the 5D case we have $\{\Gamma_{05},\Gamma_{04}\Gamma^6\}=0$.
To solve this we write $P_5=M\sin\theta$ and $Z^6_4 = M\cos \theta$ so that we have the  operator  $M(1+P )$ where $P = \sin\theta\Gamma_{05}-\cos\theta \Gamma_{04}\Gamma^6$ and  $P_\pm^2=1$. So in total we find a $\frac{1}{2}$-BPS state in agreement with the 5D analysis above.

From the six-dimensional point of view it might seem odd that by adding momentum along $x^5$ we break an extra half of the supersymmetry in the case of a self-dual string that is extended along $x^5$ but not when it is extended along some other direction. After all, by a Lorentz transformation one can go back to the rest frame where there is no momentum. However it is important to note that the central charges are not Lorentz scalars. In the case of a self-dual string extended along $x^4$, the corresponding  $Z^I_{4}$ charge is left invariant by the boost required to go back to the rest frame. However a self-dual string which is extended along $x^5$ and carries non-zero $Z^I_{5}$ charge picks-up a $Z^I_{0}$ charge after the boost. Thus in the new frame one is simply not looking at a self-dual string but some other bound state.

In addition note that the ground state of a string extended along $x^5$ has Poincar\'e symmetry in the $x^0,x^5$ plane. Thus it is invariant under a boost in the $x^0,x^5$ plane that could give it non-vanishing $x^5$ momentum. This means that the states that the dyonic instantons describe should be thought of as $\frac{1}{2}$-BPS excitations of the self-dual string that carry some $x^5$ momentum. Whereas the monopole strings that are stretched along $x^4$ are ground-states of the self-dual string but carry $x^5$ momentum. One can therefore also look for $\frac{1}{2}$-BPS excited states of these strings. These states should agree with the spectrum of dyonic instantons when the transverse space is ${\mathbb R}^3\times S^1$ rather than ${\mathbb R}^4$. In fact this is trivially true since the instanton equations on ${\mathbb R}^3\times S^1$ (and with no dependence on the circle)  are just the monopole equations. In both cases we are effectively looking at $\frac{1}{4}$-BPS monopoles in 4D super-Yang-Mills.

\section{Other States}\label{other}

In the previous two sections we showed that 5D super-Yang-Mills contains BPS solutions corresponding to the full KK tower of self-dual string states and thus `knows' about the extra dimension associated to the M5-brane. There are of course other BPS states and it is natural to consider whether or not these also admit a KK tower. However we will see that the answer is no.

One can ask if there are chargeless instanton states with a non-vanishing vacuum expectation value for $X^6$. However
this cannot be the case since
\begin{eqnarray}
\int  \tr( D_iX^6 D_iX^6) d^4x =\oint \tr( X^6 D_iX^6)dS_i  =   \; Q_E\ .
\end{eqnarray}
Thus a vanishing electric charge would imply $D_iX^6=0$ and this in turn implies $[F_{ij},X^6]=0$. Since this is not possible we conclude that the only smooth instanton particle-like solutions in a vacuum with $\langle X^6\rangle\ne 0$ carry electric charge.

It is natural to ask what BPS states carry $Z^I_{0}$ charge. From the $(2,0)$ algebra we see that the supersymmetries are simply eigenstates of $\Gamma^6 $ or, equivalently, $\Gamma^{78910}\epsilon = \epsilon$.  Furthermore, such states do not carry momentum along $x^5$ if they are to preserve half of the supersymmetry, and thus should appear in five-dimensional super-Yang-Mills. It is not hard to see that they are provided by
\begin{equation}
  [X^I,X^J] =-\frac{1}{2} \epsilon^{6IJKL}[X^K,X^L]\ ,\qquad A_\mu=\partial_\mu X^I=0\;,
\end{equation}
\ie\ instanton-like solutions in the transverse space. Note that these equations imply $[X^6,X^I]=0$. As a result, such solutions do not exist for gauge group $\SU(2)$ in the broken phase where $X^6\ne 0$. Indeed they only exist when the unbroken gauge group is non-Abelian, which we do not consider here. Furthermore the M-theory interpretation of such a solution is not clear since there is no Abelian analogue.

We could also try to look for KK states that carry $Z_{ijk}^{IJ}$ charges.  These correspond to intersections of the M5-brane with another M5-brane in the $x^i,x^j,x^k, x^I,x^J$ plane and result in a 3-brane solution on the worldvolume of the original M5-brane \cite{Howe:1997et}. However there is an important distinction with the case of M2-branes and self-dual strings: an M2-brane can end on the M5 and deposit a charge but M5-branes cannot end on other M5's. Thus it is not clear to what extent an M5-brane intersecting with another M5-brane leads to a fundamental state in the theory and it has been associated to a sort-of D-brane in the $(2,0)$ string theory \cite{Howe:1997et}.

Indeed if we look for solutions we find that they cannot have a non-zero instanton number. Let us look at the two possible cases. Firstly consider an M5-brane that lies in the $x^1,x^2,x^3,x^6,x^7$ plane, on top of our original M5 along $x^1, x^2, x^3, x^4, x^5$. The resulting 3-brane solution should have Poincar\'e symmetry along $x^1,x^2,x^3$. However this means that in the reduced theory $F_{ab}=0$, $a,b = 1,2,3$ and hence $F\wedge F = 0$. The other possible configuration is when the intersecting M5-brane lies in the $x^3,x^4,x^5,x^6,x^7$ plane, \ie ~it also extends along the M-theory direction. Now the Poincar\'e symmetry is along $x^3,x^4$. This allows for $F_{12}$ and $F_{34}$ to be non-vanishing, so that $F\wedge F$ could also be non-vanishing. However it also requires $F_{13}=F_{23}=F_{14}=F_{24}=0$ and $D_3X^6=D_4X^6=D_3X^7=D_4X^7=0$. By additionally imposing that there is translational invariance along $x^3,x^4$ one finds the conditions $[A_3,X^6]=[A_4,X^6]=[A_3,X^7]=[A_4,X^7]=0$. At least for $\SU(2)$ these conditions imply $[A_3,A_4]=iF_{34}=0$ and hence $F\wedge F=0$.

\section{A Higgs Mechanism for tensor non-Abelian multiplets}\label{Higgs}

To summarise our results so far, we saw that we are able to identify instanton states that correspond to a KK tower of charged string states. These are analogues of $W^\pm$-Bosons in the gauge theory. In M-theory these correspond to self-dual string states that carry some non-zero central charge $Z^I_m$, $m\ne 0$. We also saw that upon quantisation the dyonic instanton moduli space  admits a $\frac{1}{2}$-BPS uncharged ground state. These give a  tower whose Bosonic fields are a self-dual tensor and 5 scalars but no vectors.

This is in agreement with the fact that   the $(2,0)$ theory is not a Yang-Mills gauge theory based on a non-Abelian vector potential.  Rather it contains a non-Abelian 2-form potential. Little is known about such theories and it is certainly possible that the Higgs mechanism is different.
Indeed, already at the level of the Abelian theory, we can see that there is no KK tower of photon vector modes   and in fact the uncharged multiplet we found in \eqref{tensormult} was predicted to arise in  \cite{Hull:2000cf}.

In particular, the linearised  equations of motion for an Abelian $(2,0)$ tensor multiplet   are\footnote{These were first derived in linearised form in \cite{Howe:1983fr} and then the full non-linear equations in \cite{Howe:1996yn} in superspace, with their component form given  in \cite{Howe:1997fb}.}
\begin{equation}
H_{mnp} = \frac{1}{3!}\varepsilon_{mnpqrs}H^{qrs}\;,\qquad \partial_{[m}H_{npq]}=0\;,\qquad \partial_m\partial^m X^I=\Gamma^m\partial_m\Psi=0\;.
\end{equation}
Let us focus on the 3-form which we express as $H = 3\partial_{[m}B_{np]}$. This has the gauge invariance $B_{mn}\to B_{mn} + \partial_m\lambda_n - \partial_n\lambda_m$.  Let us also suppose that the $x^5$ direction is compact with period $2\pi R_5$ and expand the fields in terms of their Fourier modes:
\begin{equation}
B_{mn} = \sum_k e^{ikx^5/R_5} B^{(k)}_{mn}\;,
\end{equation}
so that $(B^{(k)}_{mn})^* = B^{(-k)}_{mn}$.

For the zero-modes $k=0$ we find that the self-duality simply relates $H^{(0)}_{\mu\nu\lambda}=3\partial_{[\mu}B^{(0)}_{\nu\lambda]}$ to $H^{(0)}_{\mu\nu 5} = 2\partial_{[\mu} B^{(0)}_{\nu] 5}$. Thus it is sufficient to just consider the Abelian gauge field $A_\mu = B^{(0)}_{\mu 5}$ with equation of motion $\partial^\mu F_{\mu\nu}=0$, where $F_{\mu\nu} = H^{(0)}_{\mu\nu 5}$.  Since one also has the residual gauge transformations
$A_\mu \to A_\mu + \partial_\mu\lambda^{(0)}_5$, we recover a familiar massless Maxwell field.

But let us now turn to the massive modes. Here things are quite different. Under a gauge transformation we have
\begin{equation}
B^{(k)}_{\mu\nu} \to B^{(k)}_{\mu\nu}+\partial_\mu\lambda^{(k)}_\nu-\partial_\nu\lambda^{(k)}_\mu\;, \qquad
B^{(k)}_{\mu 5} \to B^{(k)}_{\mu 5}+\partial_\mu\lambda^{(k)}_5-i\frac{k}{R}\lambda^{(k)}_\mu
\end{equation}
and we can completely gauge away $B^{(k)}_{\mu 5}$ by a choice of $\lambda^{(k)}_\mu$. As a result we are simply left with the self-duality condition
\begin{equation}
B^{(k)}_{\mu\nu} = - \frac{iR_5}{2k}\varepsilon_{\mu\nu\lambda\rho\sigma}\partial^\lambda B^{(k)\rho\sigma}\;,
\end{equation}
which implies that $B^{(k)}_{\mu\nu}$ satisfies
\begin{equation}
\partial_\lambda\partial^\lambda B^{(k)}_{\mu\nu}  - \frac{k^2}{R^2_5}B^{(k)}_{\mu\nu}=0
 \end{equation}
 so that $B^{(k)}_{\mu\nu}$ is indeed a massive field.\footnote{This was also discussed in \cite{Lee:2000kc} and \cite{Hull:2000cf}.}
In addition we obtain 5 massive KK scalars from the scalars of the six-dimensional theory. Note that in this Abelian limit the R-symmetry is unbroken from $\SO(5)$, but in the non-Abelian theory of interest it will be broken to $\SO(4)$. Thus the instanton spectrum that we obtain above agrees precisely with the what is expected from the reduction of the $(2,0)$ theory.\footnote{See also  \cite{Hull:2000cf}.}

The non-Abelian version of this mechanism would be interesting to understand better. In the familiar 6D Higgs mechanism a non-Abelian gauge field eats a single scalar degree of freedom to become massive. In doing so, the number of degrees of freedom for the gauge field goes from $4$ to $5$. However for a massive 2-form the counting is different. A massless self-dual   2-form  in six dimensions has 3 physical degrees of freedom whereas a massive one has 5. In this case it cannot become massive by eating a single scalar field.

\section{Conclusion}\label{conclusion}

In this paper we have matched the KK spectrum for charged self-dual string states of the $(2,0)$ theory compactified on $S^1$ with instanton solitons of five-dimensional super-Yang-Mills, in the broken phase when all the branes are slightly separated.   It therefore seems reasonable to conjecture that 5D SYM is precisely the $(2,0)$ CFT on $S^1$, for any value of the radius. We note that in a recent paper an attempt was made to construct non-Abelian $(2,0)$ theories in six dimensions using 3-algebras \cite{Lambert:2010wm}. However it was found that the non-Abelian modes where constrained to propagate in five-dimensions. This is consistent with what we find here. In particular, one should not be able to have both six-dimensional momentum and non-Abelian instantons.

Since the M5-brane CFT is finite it follows that it remains finite once compactified. Given our claim that this theory is precisely five-dimensional super-Yang-Mills, this statement strongly suggests that the latter is also finite, despite being power-counting non-renormalisable. Indeed, following the remarkable progress and results in the divergences of maximally supersymmetric field theories, it is now known that five-dimensional super-Yang-Mills  is finite up to 5 loops \cite{Bern:1998ug,Bern:1999vk,Bern:2000mf}.  In addition, holographic renormalisation of D$p$-brane theories behaves in qualitatively the same  way for  $p= 1,2,3,4$ \cite{Kanitscheider:2008kd}. Since these theories are finite for $p\le 3$ this points towards a finite or relatively simple UV structure of 5D super-Yang-Mills. Some  other recent work on the quantum properties of this theory in light-cone superspace has appeared  in \cite{Brink:2010ti}.

It should be emphasised that the relationship between the $(2,0)$ and the super-Yang-Mills finiteness may not be so simple, since the 6D CFT contains momentum states which are non-perturbative from the point of view of the five-dimensional theory. Moreover, as stressed in \cite{Collie:2009iz}, although the theory might know about the existence of the KK states it may not know enough about their dynamics. In particular, from the six-dimensional point of view, once above the KK scale any scattering of states must create states with momentum in the extra dimension. Thus, in terms of five-dimensional super-Yang-Mills, once above the energy scale $g_{YM}^{-2}$ one must produce instanton-anti-instanton pairs. An alternative, or perhaps complementary, possibility is that there is some kind of variation of the `classicalisation' process of \cite{Dvali:2010jz}.

In this paper, we have concentrated on the case of two M5-branes emerging in the UV from an $\SU(2)$ gauge group on two D4-branes of type IIA. However, it is clearly of interest to generalise these results to an arbitrary number of branes with gauge group $\SU(N)$. In particular it is not clear that all the required solutions that we have constructed here have appropriate analogues for the case of $N>2$. It would also be interesting to relate our discussion to those of \cite{Dijkgraaf:1996hk,Henningson:2009dz}.

\section*{Acknowledgements}

We would like to thank Frederik Denef, Lance Dixon, Greg Moore, Kostas Skenderis, Julian Sonner,  Andy Strominger, Marika Taylor, Alessandro Tomasiello and David Tong for various discussions and comments. We would also like to thank Mike Douglas for sending us an advance copy of \cite{Douglas:2010iu} and Chris Hull for pointing-out the relevance of \cite{Hull:2000cf}. C.P. would like to thank Harvard and Rutgers University for hospitality during the course of this work and is supported by the STFC rolling grant ST/G000395/1.

\bibliographystyle{utphys}
\bibliography{finiteD4}

\providecommand{\href}[2]{#2}\begingroup\raggedright\begin{thebibliography}{10}

\bibitem{Rozali:1997cb}
M.~Rozali, ``{Matrix theory and U-duality in seven dimensions},''
  \href{http://dx.doi.org/10.1016/S0370-2693(97)00361-4}{{\em Phys. Lett.} {\bf
  B400} (1997)  260--264},
\href{http://arxiv.org/abs/hep-th/9702136}{{\tt arXiv:hep-th/9702136}}.

\bibitem{Berkooz:1997cq}
M.~Berkooz, M.~Rozali, and N.~Seiberg, ``{Matrix description of M theory on
  T**4 and T**5},'' \href{http://dx.doi.org/10.1016/S0370-2693(97)00800-9}{{\em
  Phys. Lett.} {\bf B408} (1997)  105--110},
\href{http://arxiv.org/abs/hep-th/9704089}{{\tt arXiv:hep-th/9704089}}.

\bibitem{Seiberg:1997ax}
N.~Seiberg, ``{Notes on theories with 16 supercharges},''
  \href{http://dx.doi.org/10.1016/S0920-5632(98)00128-5}{{\em Nucl. Phys. Proc.
  Suppl.} {\bf 67} (1998)  158--171},
\href{http://arxiv.org/abs/hep-th/9705117}{{\tt arXiv:hep-th/9705117}}.

\bibitem{Jackiw:1975fn}
R.~Jackiw and C.~Rebbi, ``{Solitons with Fermion Number 1/2},''
\href{http://dx.doi.org/10.1103/PhysRevD.13.3398}{{\em Phys. Rev.} {\bf D13}
  (1976)  3398--3409}.

\bibitem{Osborn:1979tq}
H.~Osborn, ``{Topological Charges for N=4 Supersymmetric Gauge Theories and
  Monopoles of Spin 1},''
\href{http://dx.doi.org/10.1016/0370-2693(79)91118-3}{{\em Phys. Lett.} {\bf
  B83} (1979)  321}.

\bibitem{Klebanov:1996un}
I.~R. Klebanov and A.~A. Tseytlin, ``{Entropy of Near-Extremal Black
  p-branes},'' \href{http://dx.doi.org/10.1016/0550-3213(96)00295-7}{{\em Nucl.
  Phys.} {\bf B475} (1996)  164--178},
\href{http://arxiv.org/abs/hep-th/9604089}{{\tt arXiv:hep-th/9604089}}.

\bibitem{Strominger:1995ac}
A.~Strominger, ``{Open p-branes},''
  \href{http://dx.doi.org/10.1016/0370-2693(96)00712-5}{{\em Phys.Lett.} {\bf
  B383} (1996)  44--47}, \href{http://arxiv.org/abs/hep-th/9512059}{{\tt
  arXiv:hep-th/9512059 [hep-th]}}.

\bibitem{Howe:1997ue}
P.~S. Howe, N.~Lambert, and P.~C. West, ``{The Selfdual string soliton},''
  \href{http://dx.doi.org/10.1016/S0550-3213(97)00750-5}{{\em Nucl.Phys.} {\bf
  B515} (1998)  203--216}, \href{http://arxiv.org/abs/hep-th/9709014}{{\tt
  arXiv:hep-th/9709014 [hep-th]}}.

\bibitem{Witten:1995zh}
E.~Witten, ``{Some comments on string dynamics},''
\href{http://arxiv.org/abs/hep-th/9507121}{{\tt arXiv:hep-th/9507121}}.

\bibitem{Lambert:1999ua}
N.~D. Lambert and D.~Tong, ``{Dyonic instantons in five-dimensional gauge
  theories},'' \href{http://dx.doi.org/10.1016/S0370-2693(99)00894-1}{{\em
  Phys. Lett.} {\bf B462} (1999)  89--94},
\href{http://arxiv.org/abs/hep-th/9907014}{{\tt arXiv:hep-th/9907014}}.

\bibitem{Douglas:2010iu}
M.~R. Douglas, ``{On D=5 super Yang-Mills theory and (2,0) theory},''
  \href{http://arxiv.org/abs/1012.2880}{{\tt arXiv:1012.2880 [hep-th]}}.

\bibitem{Howe:1997et}
P.~S. Howe, N.~Lambert, and P.~C. West, ``{The three-brane soliton of the
  M-five-brane},'' \href{http://dx.doi.org/10.1016/S0370-2693(97)01433-0}{{\em
  Phys.Lett.} {\bf B419} (1998)  79--83},
  \href{http://arxiv.org/abs/hep-th/9710033}{{\tt arXiv:hep-th/9710033
  [hep-th]}}.

\bibitem{D'Auria:2000ec}
R.~D'Auria, S.~Ferrara, M.~A. Lledo, and V.~S. Varadarajan, ``{Spinor
  algebras},'' \href{http://dx.doi.org/10.1016/S0393-0440(01)00023-7}{{\em J.
  Geom. Phys.} {\bf 40} (2001)  101--128},
\href{http://arxiv.org/abs/hep-th/0010124}{{\tt arXiv:hep-th/0010124}}.

\bibitem{Hull:2000cf}
C.~Hull, ``{BPS supermultiplets in five-dimensions},'' {\em JHEP} {\bf 0006}
  (2000)  019, \href{http://arxiv.org/abs/hep-th/0004086}{{\tt
  arXiv:hep-th/0004086 [hep-th]}}.

\bibitem{BlancoPillado:2006qu}
J.~J. Blanco-Pillado and M.~Redi, ``{Supersymmetric rings in field theory},''
  \href{http://dx.doi.org/10.1088/1126-6708/2006/09/019}{{\em JHEP} {\bf 0609}
  (2006)  019}, \href{http://arxiv.org/abs/hep-th/0604180}{{\tt
  arXiv:hep-th/0604180 [hep-th]}}.

\bibitem{Howe:1983fr}
P.~S. Howe, Sierra, and P.~Townsend, ``{Supersymmetry in Six Dimensions},''
  {\em Nucl. Phys} {\bf B221} (1983)  331.

\bibitem{Howe:1996yn}
P.~S. Howe and E.~Sezgin, ``{D = 11, p = 5},''
  \href{http://dx.doi.org/10.1016/S0370-2693(96)01672-3}{{\em Phys.Lett.} {\bf
  B394} (1997)  62--66}, \href{http://arxiv.org/abs/hep-th/9611008}{{\tt
  arXiv:hep-th/9611008 [hep-th]}}.

\bibitem{Howe:1997fb}
P.~S. Howe, E.~Sezgin, and P.~C. West, ``{Covariant field equations of the M
  theory five-brane},''
  \href{http://dx.doi.org/10.1016/S0370-2693(97)00257-8}{{\em Phys.Lett.} {\bf
  B399} (1997)  49--59}, \href{http://arxiv.org/abs/hep-th/9702008}{{\tt
  arXiv:hep-th/9702008 [hep-th]}}.

\bibitem{Lee:2000kc}
K.-M. Lee and J.-H. Park, ``{5-D actions for 6-D selfdual tensor field
  theory},'' \href{http://dx.doi.org/10.1103/PhysRevD.64.105006}{{\em
  Phys.Rev.} {\bf D64} (2001)  105006},
  \href{http://arxiv.org/abs/hep-th/0008103}{{\tt arXiv:hep-th/0008103
  [hep-th]}}.

\bibitem{Lambert:2010wm}
N.~Lambert and C.~Papageorgakis, ``{Nonabelian (2,0) Tensor Multiplets and
  3-algebras},'' \href{http://dx.doi.org/10.1007/JHEP08(2010)083}{{\em JHEP}
  {\bf 1008} (2010)  083}, \href{http://arxiv.org/abs/1007.2982}{{\tt
  arXiv:1007.2982 [hep-th]}}.

\bibitem{Bern:1998ug}
Z.~Bern, L.~J. Dixon, D.~C. Dunbar, M.~Perelstein, and J.~S. Rozowsky, ``{On
  the relationship between Yang-Mills theory and gravity and its implication
  for ultraviolet divergences},''
  \href{http://dx.doi.org/10.1016/S0550-3213(98)00420-9}{{\em Nucl. Phys.} {\bf
  B530} (1998)  401--456},
\href{http://arxiv.org/abs/hep-th/9802162}{{\tt arXiv:hep-th/9802162}}.

\bibitem{Bern:1999vk}
Z.~Bern, L.~J. Dixon, M.~Perelstein, D.~C. Dunbar, and J.~S. Rozowsky,
  ``{Perturbative relations between gravity and gauge theory},''
  \href{http://dx.doi.org/10.1088/0264-9381/17/5/307}{{\em Class. Quant. Grav.}
  {\bf 17} (2000)  979--988},
\href{http://arxiv.org/abs/hep-th/9911194}{{\tt arXiv:hep-th/9911194}}.

\bibitem{Bern:2000mf}
Z.~Bern {\em et al.}, ``{On perturbative gravity and gauge theory},''
  \href{http://dx.doi.org/10.1016/S0920-5632(00)00768-4}{{\em Nucl. Phys. Proc.
  Suppl.} {\bf 88} (2000)  194--203},
\href{http://arxiv.org/abs/hep-th/0002078}{{\tt arXiv:hep-th/0002078}}.

\bibitem{Kanitscheider:2008kd}
I.~Kanitscheider, K.~Skenderis, and M.~Taylor, ``{Precision holography for
  non-conformal branes},''
  \href{http://dx.doi.org/10.1088/1126-6708/2008/09/094}{{\em JHEP} {\bf 0809}
  (2008)  094}, \href{http://arxiv.org/abs/0807.3324}{{\tt arXiv:0807.3324
  [hep-th]}}.

\bibitem{Brink:2010ti}
L.~Brink and S.~S. Kim, ``{Maximally Supersymmetric Yang-Mills in five
  dimensions in light-cone superspace,},''
\href{http://arxiv.org/abs/1011.5817}{{\tt arXiv:1011.5817 [hep-th]}}.

\bibitem{Collie:2009iz}
B.~Collie and D.~Tong, ``{The Partonic Nature of Instantons},''
  \href{http://dx.doi.org/10.1088/1126-6708/2009/08/006}{{\em JHEP} {\bf 08}
  (2009)  006},
\href{http://arxiv.org/abs/0905.2267}{{\tt arXiv:0905.2267 [hep-th]}}.

\bibitem{Dvali:2010jz}
G.~Dvali, G.~F. Giudice, C.~Gomez, and A.~Kehagias, ``{UV-Completion by
  Classicalization},'' \href{http://arxiv.org/abs/1010.1415}{{\tt
  arXiv:1010.1415 [hep-ph]}}.

\bibitem{Dijkgraaf:1996hk}
R.~Dijkgraaf, E.~P. Verlinde, and H.~L. Verlinde, ``{BPS quantization of the
  five-brane},'' \href{http://dx.doi.org/10.1016/S0550-3213(96)00639-6}{{\em
  Nucl.Phys.} {\bf B486} (1997)  89--113},
  \href{http://arxiv.org/abs/hep-th/9604055}{{\tt arXiv:hep-th/9604055
  [hep-th]}}.

\bibitem{Henningson:2009dz}
M.~Henningson, ``{BPS states in (2,0) theory on R x T**5},''
  \href{http://dx.doi.org/10.1088/1126-6708/2009/03/021}{{\em JHEP} {\bf 0903}
  (2009)  021}, \href{http://arxiv.org/abs/0901.0785}{{\tt arXiv:0901.0785
  [hep-th]}}.

\end{thebibliography}\endgroup

\end{document}